\documentclass[12pt]{article}

\usepackage[ansinew]{inputenc}
\usepackage[bookmarks=false]{hyperref}
\usepackage[english]{babel}
\usepackage[right]{eurosym}
\usepackage[T1]{fontenc}
\usepackage{amsfonts}
\usepackage{amsmath}
\usepackage{amssymb}
\usepackage{amsthm}
\usepackage{array}
\usepackage{arydshln}
\usepackage{booktabs}
\usepackage[singlelinecheck=false]{caption}
\usepackage{color}
\usepackage{eurosym}
\usepackage{float}
\usepackage{floatpag}
\usepackage{geometry}
\usepackage{graphicx}
\usepackage{hyphenat}
\usepackage{lmodern}
\usepackage{multirow}
\usepackage{natbib}
\usepackage{rotating}
\usepackage{setspace}
\usepackage{siunitx}
\usepackage{subfig}
\usepackage{tabularx}
\usepackage{tikz}
\usepackage{pdfpages}
\usepackage{pbox}
\usepackage{epstopdf}

\newcommand{\E}{\mathbf{E}}

\newcommand{\R}{{\mathbb R}}

\newcommand{\x}{\mathbf x}
\newcommand{\X}{\mathbf X}

\DeclareMathOperator*{\argmin}{arg\,min}

\newcolumntype{C}[1]{>{\centering\arraybackslash}m{#1}}
\newcolumntype{L}[1]{>{\raggedright\arraybackslash}m{#1}}
\newcolumntype{R}[1]{>{\raggedleft\arraybackslash}m{#1}}

\title{\vspace{-.6cm}Robust Productivity Analysis: An application to German FADN data}

\author{Mathias Kloss\thanks{Leibniz Institute of Agricultural Development in Transition Economies (IAMO), Theodor-Lieser-Stra{\ss}e 2, D-06120, Halle (Saale), Germany} \and Thomas Kirschstein\thanks{Martin-Luther-University Halle-Wittenberg, Gro{\ss}e Steinstra{\ss}e 73, D-06099 Halle (Saale), Germany} \and Steffen Liebscher\footnotemark[2] \and Martin Petrick\footnotemark[1]}

\begin{document}


\maketitle

\begin{abstract}
Sources of bias in empirical studies can be separated in those coming from the modelling domain (e.g. multicollinearity) and those coming from outliers. We propose a two-step approach to counter both issues. First, by decontaminating data with a multivariate outlier detection procedure and second, by consistently estimating parameters of the production function. We apply this approach to a panel of German field crop data. Results show that the decontamination procedure detects multivariate outliers. In general, multivariate outlier control delivers more reasonable results with a higher precision in the estimation of some parameters and seems to mitigate the effects of multicollinearity. \newpage
\end{abstract}

\section{Introduction and Background}
\label{intro}

In assessing factor productivity, biases can mainly emerge from two different sources. The first source is with regard to the classical identification problem  of production function parameters. Inputs are usually subject to the farmers decision making process. Hence, there is an endogeneity problem due to the endogenous nature of input choice \citep{Griliches1998}. In addition, due to lack of variation in factor utilisation across firms, standard estimation procedures like OLS and fixed effects are not able to extract the information necessary for the separate identification of the different output elasticities. This leads to a collinearity problem \citep{Ackerberg2007}.   Recently, there has been new insight into these two issues and promising approaches have been developed to mitigate them \citep[cf.][]{Olley1996,Blundell2000, Levinsohn2003,Wooldridge2009, Gandhi2011}. However, even in case that the statistical identification is secured, there  might be  sources of bias emerging from the data itself. This usually happens if there are \textit{outliers} in the data set, which is an ubiquitous feature of many real world data sets, and hence an issue in many empirical applications. If outliers are not dealt with, estimators can be obscured arbitrarily by the presence of as little as one single outlier in the data set. Moreover, in terms of estimating factor productivity accounting for outliers ensures that the assumption of a homogenous production technology is maintained. In this paper, we focus on this latter source of bias while treating the production function estimator which controls for the identification problems mentioned as given.

	Outliers occur e.g.\,due to measurement errors, variations in the data generating process, or misreporting. In general, other than by being very pragmatic and not performing any outlier decontamination, there are two concurrent views on how to define and, consequently, identify outliers.\footnote{Not performing any outlier control is a common practice (see section 2).} On the one hand, there are methods which assume a rigorous statistical model e.g.\,in a sense that outliers are thought of as data coming from a different distribution (from another than the one the researcher is actually interested in). The whole data can then be understood as a mixture of two (or more) distributions, where the target distribution should comprise the majority of the probability mass (if, in contrast, the outliers would constitute the majority, it would not make much sense to speak of outliers at all. In such cases the outliers should rather be considered as the distribution of interest). Methods falling into that category commonly aim at estimating some features of the target distribution while outlier detection is usually not their primary objective (it is more of an inherent by-product). If outliers are identified, they are identified with that precise model in mind, meaning that the methods in that category cannot be reasonably applied when the underlying (distributional) assumptions are not fulfilled (at least no useful results can be expected in that case). In this category, a bunch of robust estimators for various models can be found, see e.g.\,\cite{Rousseeuw1987,Barnett2000,Hampel2005,Maronna2006,Huber2009} for an overview of such methods.
	
	On the other hand, there are methods which follow a more  sample-oriented view on how to define an outlier. Universally, all of these methods interpret outliers as observations that differ from the target observations. Obviously, there exist numerous proposals on how that difference can actually be quantified. A large amount of these methods is based on some type of distance, either between any two observations or with respect to some reference point, but there are also other approaches e.g. based on depth or on the empirical density. Representatives can mainly be found in the computer science and data mining literature, see e.g.\,\cite{Chandola2009} and references therein as a starting point. With these methods, the primary goal is indeed the identification of outliers. An additional analysis might still be carried out on the identified non-outliers, though. The biggest advantage of methods falling into that category is that they can generally be applied without being limited to situations where certain distributional assumptions are fulfilled.
	
	Generally, as we discovered by evaluating a sample of empirical economics papers, if a decontamination is conducted almost always univariate methods focusing on one variable of a more complex multivariate model are employed prior to the follow-up analysis. However, such approaches neglect the models multivariate nature.
	
	


In this paper we propose a robust two-stage approach for estimating the production function for field crop farms in East and West Germany based on the "Farm Accountancy Data Network" (FADN) data set. First, by performing the outlier decontamination, and second by estimating the parameters of the production function. While the literature on outlier detection methods is vast \citep{Maronna2006} and other methods might be feasible as well (e.g. cluster analysis), we resort to a multivariate decontamination procedure due to \cite{Kirschstein2013} for identifying the outliers. This method has already proven its effectiveness in an application to determine unsuccessful warship designs \citep{Liebscher2012a}. Moreover, it is a non-parametric approach (not requiring special distributional assumptions) and it offers computational advantages (especially for large data sets as the one analyzed here) which makes it our preferred choice. After decontamination, we estimate the parameters of the production function by the \cite{Wooldridge2009} instrumental variable estimator to account for endogeneity as well as collinearity issues. While we consider the estimation of production functions, the insights should also be of relevance to other model driven empirical applications in which the data is contaminated by outliers as well as those in which instrumental variable estimation is performed.

The paper proceeds as follows. In section 2 we discuss the going practice in outlier treatment in empirical economics.  Next, we present the methodological background. Firstly, by discussing our approach to outlier identification. Secondly, by presenting the methods employed for production function estimation. At this stage, we also illustrate by example the consequences to the production function estimation in presence of outliers in the data. Section 4 discusses the FADN data. We move on with a presentation of our results. Section 6 concludes the paper.

\section{Going practice in empirical economics}

In order to get a general idea how outliers have been treated within the field of economics, we surveyed studies from two additional data sources that have approximately the same importance as the "Farm Accountancy Data Network" (FADN), our data source, in agricultural economics. For the FADN data, we reviewed studies from two recent research projects funded by the European Commission under FP7 which heavily employed this data set: "Farm Accountancy Cost Estimation and Policy Analysis of European  Agriculture" (FACEPA) and "Factor Markets". For the field of development economics we reviewed Working Papers from the World Bank's "Living Standard Measurement Survey" (LSMS) studies while for the field of general economics we analyzed studies employing data from the "German Socio-Economic Panel" (GSOEP). The GSOEP's "SOEPapers" series is a collection of such work.  In \autoref{tab:goingpr} we summarize the going practice of outlier treatment for a sample of empirical work from  these three data sources.


Generally, the three data sources leave the assessment as well as handling of outliers to the researcher \citep{Grosh1996, Haisken-DeNew2005, EuropeanCommission2010}. \citet[][p. 125]{Grosh1996} for the LSMS are explicitly open about this view by explaining that "further treatment of these problems should be left to analysts, since there is no universally acceptable solution to these problems". This argument opens up many possibilities for researchers to deal with outliers.

In empirical economic literature emerging from these sources, a large group of authors assess outliers by visual observation - which is usually only feasible for smaller data sets and two dimensions - and contextual reasoning. Based on these modes of operation they drop implausible cases \citep[e.g.][]{Deaton1981, Deaton1988, Crosetto2012, Obschonka2013, Oltmanns2014, Auer2014}. Other approaches involve the transformation of variables (e.g. logarithmization), the application of influence measures or censoring of extreme values \citep[cf.][]{Schneck2011, Olper2014, Liverpool-Tasie2015}. Some authors remove outliers without stating their method of detection \citep[e.g.][]{Bauernschuster2011, Ciaian2011, Lang2012, Kemptner2013, Arnold2014}.

More structured approaches apply a two-step procedure by first identifying and removing extreme observations for a target variable (univariate outliers) prior to the desired analysis. The approach most commonly used among all empirical data sources is 'trimming', either by removing some percentage of the data (usually 1\% or 5\%) at the top and/or bottom of the distribution of a single univariate measure central to the analysis or by applying a quantile based rule, e.g. upper/lower quartile $ \pm s \cdot IQR$ with $s$ being some scaling factor and $IQR$ the interquartile range \citep[e.g.][]{Pfeiffer2011, Zibrowius2012, Schmitt2013, Sorgner2013, Guastella2013, Oseni2014, Murphy2015, Avdic2015, Backiny-Yetna2015}. However, univariate outlier identification approaches neglect the multivariate data's nature.

Most strikingly, the group of papers that does not mention any outlier control is by far the largest \citep[e.g.][]{Pitt1995, Alderman1998, Bauer2011, Carletto2011, Headey2012, Dustmann2015, Schurer2015}. For instance, in only 8.6\% of the sample comprising 396 studies we reviewed from the SOEPapers data base (period 2011 - 2015) an outlier control was mentioned.

Another way is to incorporate the outlier problem directly into the estimator (i.e.\,by developing or applying a robust estimator). \citet{Hubler2012} applies a simple quantile regression estimator. This way outliers are accounted for by using an appropriate estimation method. However, with the estimation of instrumental variable problems in mind, this method is not feasible since it does not account for endogeneity and collinearity issues. To develop a robust estimator capable to deal with endogeneity and collinearity issues would require extensive work which is beyond the scope of this paper.

Another two-step approach, and so far the only one to our knowledge, which accounts for outliers in a multivariate manner using FADN data has been used within the FACEPA project \citep{FACEPA2011}. Their detection procedure employs the minimum covariance determinant (MCD) algorithm due to \citet{Rousseeuw1985c}. However, this approach imposes strict distributional assumptions on the data. Furthermore, their estimation procedure does not account for farm specific heterogeneity.

In summary, in the majority of studies no outlier control is mentioned. In those studies that do mention a decontamination procedure univariate methods prevail.

\begin{table}
\caption{Outlier treatment in empirical economics}\label{tab:goingpr}
\small
\begin{tabular}{l|c|c|c}
\hline
 & FADN & LSMS & GSOEP \\
\hline
\multirow{4}{20mm}{Sample composition} & \multirow{4}{35mm}{\centering FACEPA (2009-2011); Factor Markets (2011-2013)} & \multirow{4}{40mm}{\centering World Bank LSM (1980-2002) \& Policy Research Working Paper (2010-2015)} & \multirow{4}{40mm}{\centering SOEPapers (2011-2015)} \\
&  &  & \\
&  &  & \\
&  &  & \\
\hline
\multirow{2}{30mm}{Number of studies surveyed} &  \multirow{2}{35mm}{\centering 36}  & \multirow{2}{40mm}{\centering 129} & \multirow{2}{40mm}{\centering 396} \\
&  &  & \\
\hline
\multirow{3}{30mm}{Number of studies that deal with outliers} &  \multirow{3}{35mm}{\centering 11 ($\approx$ 30.6\%)}  & \multirow{3}{40mm}{\centering 23 ($\approx$ 17.8\%)} & \multirow{3}{40mm}{\centering 34 ($\approx$ 8.6\%)} \\
&  &  & \\
&  &  & \\
\hline
\multirow{3}{30mm}{Dominant decontamination approach} &  \multirow{3}{35mm}{\centering Univariate}  & \multirow{3}{40mm}{\centering Univariate} & \multirow{3}{40mm}{\centering Univariate} \\
&  &  & \\
&  &  & \\
\hline
\multirow{10}{30mm}{Frequently used methods and examples} & \multirow{10}{40mm}{\centering Trimming \citep{Guastella2013, Petrick2013}, No method stated \citep{Bakucs10, Ciaian2011}} & \multirow{10}{40mm}{ \centering Visual observation \citep{Deaton1981, Deaton1988}, Trimming \citep{Oseni2014, Backiny-Yetna2015}, Censoring of outliers \citep{Liverpool-Tasie2015}} & \multirow{10}{40mm}{\centering Visual observation \citep{Crosetto2012, Obschonka2013}, Trimming \citep{Pfeiffer2011, Murphy2015}, No method stated \citep{Lang2012, Arnold2014}} \\
&  &  & \\
&  &  & \\
&  &  & \\
&  &  & \\
&  &  & \\
&  &  & \\
&  &  & \\
&  &  & \\
&  &  & \\
\hline
\end{tabular} \\
Source: Authors.
\end{table}

\section{Methodology}

In order to perform an unbiased estimation, we proceed in two steps. First, we decontaminate the sample from outliers by pruning the minimum spanning tree \citep{Kirschstein2013}. After eliminating the outliers from the data, we proceed by estimating the production function using the \citet{Wooldridge2009} production function estimator. In the upcoming subsection we describe both methods in detail. We conclude this section by demonstrating the effects of outliers on production function estimation with simulated data.

\subsection{Decontamination by pruning the minimum spanning tree}
\label{sec:out_dec}

Robust statistical methods are designed to deliver unbiased estimates of measures of interest in presence of contaminated data sets. A general two-step-approach is, therefore, to identify an outlier-free subsample first. This step can be called decontamination. In turn, observations not belonging to the uncontaminated subsample are suspected to be outliers. In contrast to outlier detection approaches, this approach tries to find an outlier-free subsample instead of outliers. This way, we accept the risk to falsely discard non-outlying observations in favor of a (most likely) outlier-free subsample. 

\begin{figure}[h!]
	\caption{Illustration of pMST procedure}\label{fig_mst_exa}
	\subfloat[Minimum spanning tree \label{fig_mst_exa_1}]{\includegraphics[width=.5\textwidth]{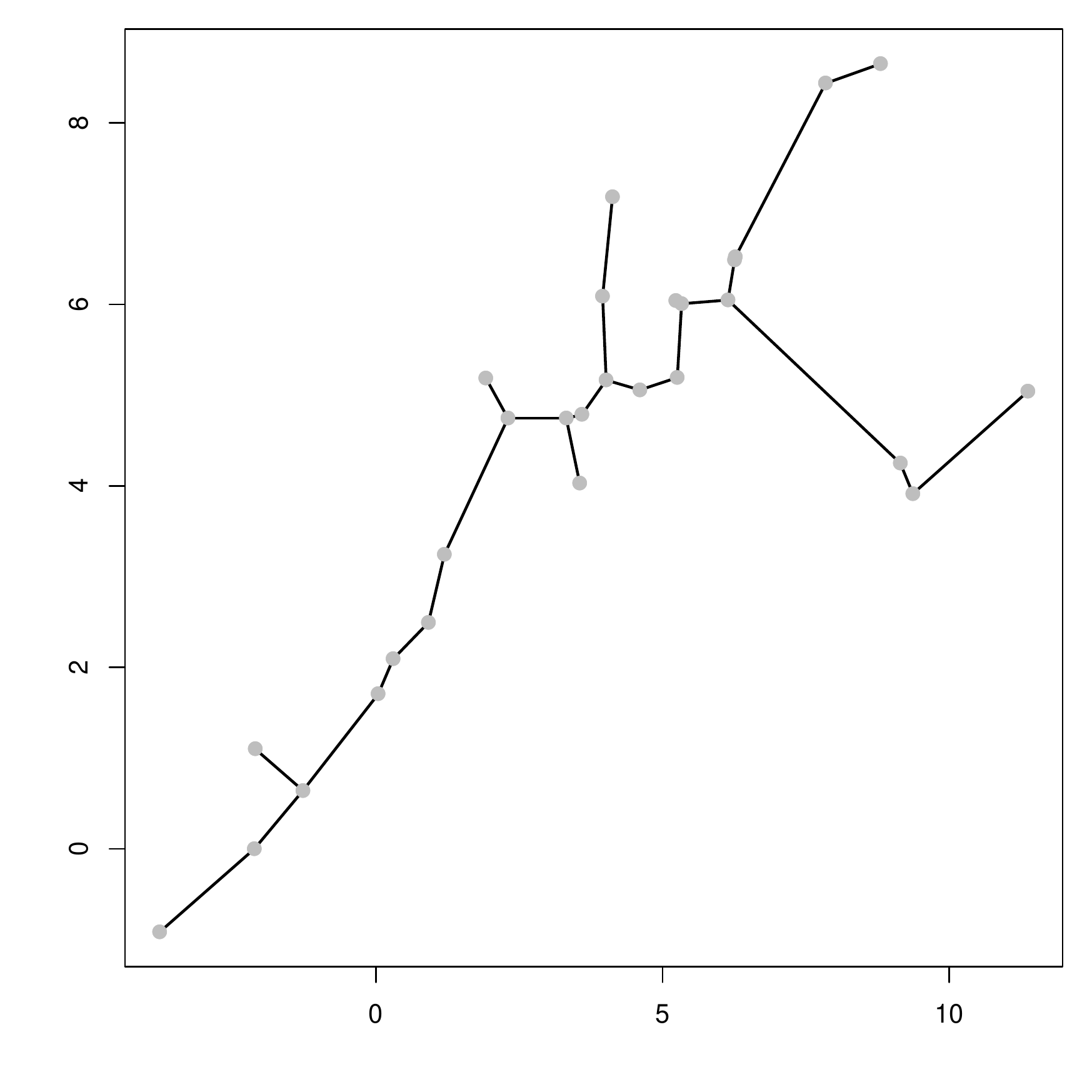}} \hfill
	\subfloat[Pruned minimum spanning tree \label{fig_mst_exa_2}]{\includegraphics[width=.5\textwidth]{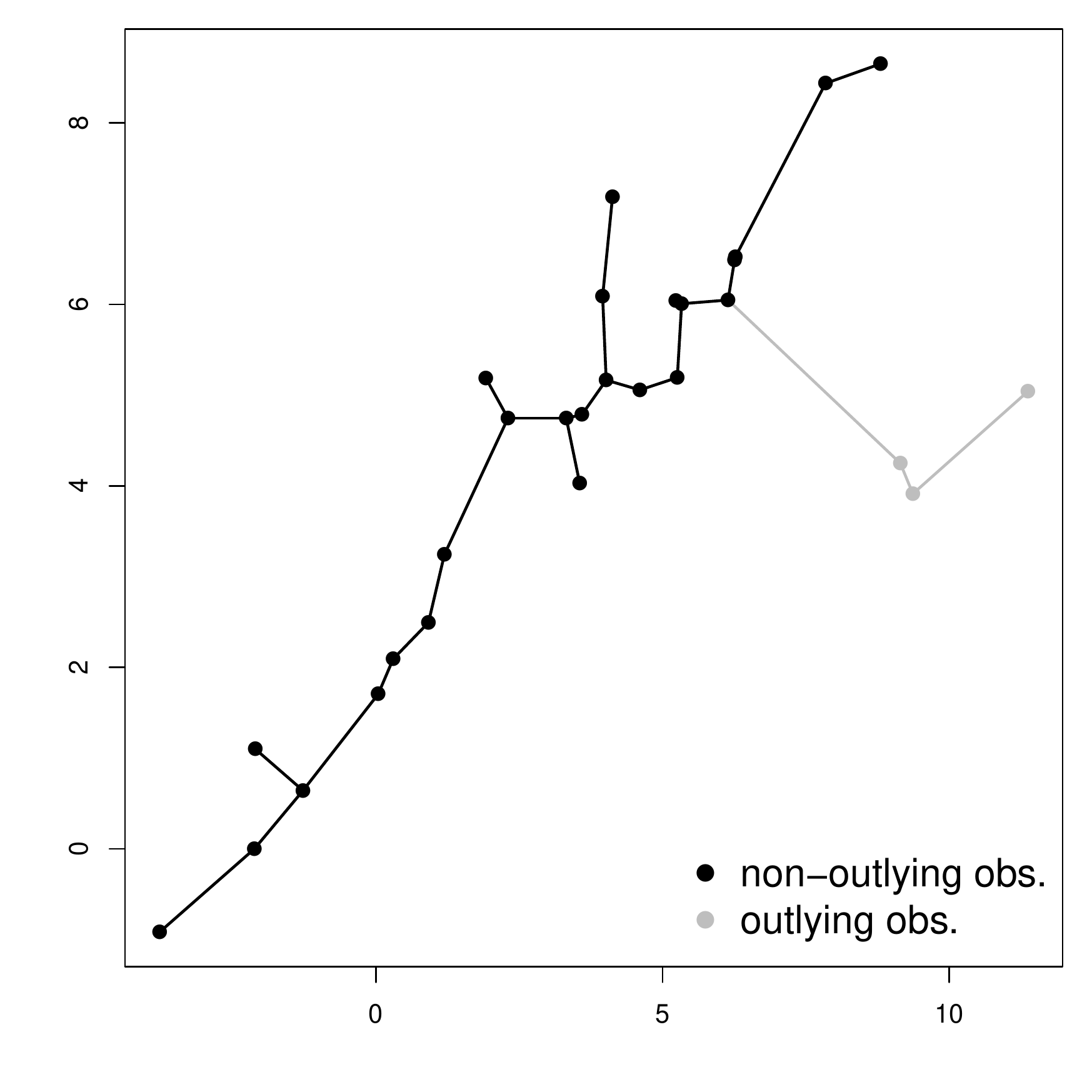}}\\
	Source: Authors based on data from \cite{Rousseeuw1987}, p.~57.
\end{figure}

A robust estimator relying on a non-parametric decontamination procedure is the pMST estimator which is a robust estimator of multivariate location and scatter \citep{Kirschstein2013}. The decontamination results from \textbf{p}runing of the so\hyp called \textbf{M}inimum \textbf{S}panning \textbf{T}ree (MST) of a data set. The idea is that each observation of a data set represents a point in Euclidean space whereby its coordinates are the observation's values in each dimension. Note that in a panel data context, an observation corresponds to an entry in the data base, e.g.~a farm's record of labour, land, materials and capital use in a certain year.\footnote{In the formal presentation we resign from using a time index to prevent a too cluttered notation.} For qualifying observations to be outlying, the pMST procedure implies that outliers are isolated with respect to similar observations. Similarity is here defined as the Euclidean distance between two observations. Similar observations can be interpreted as neighbours. In order to identify the observations' neighbourhoods, the spanning tree concept is used (see \autoref{fig_mst_exa} for a graphic illustration in two dimensions). The aim is to select a minimal set of connections between observations such that all observations are connected with each other. Among all spanning trees, the minimum spanning tree has smallest weight, i.e.~the sum of the lengths of all connections is minimal. For decontamination, the pMST procedure then iteratively deletes the longest connections in the MST until a certain threshold is reached (see below). The largest (still connected) fraction of the original MST is finally retained as the non-outlying part of the original data set.
A formal description of the algorithm follows. 

	%

 Formally, given a data set $\X$ of $n$ points in dimension $p$, i.e.\,$\X = \{ \x_1, \ldots, \x_n \} \subseteq \R^p$, and all pairwise links (edges) $\E$ (i.e.\,$\E \ni e_{ij} = \{\x_i,\x_j\}$ with $i,j=1,...,n$ and $i\neq j$), the MST is defined as the graph $\mathbf{G}=(\X,\E^*)$ connecting all points of $\X$ such that its total length is minimized, i.e.\,$\argmin\limits_{\E^*\subset\E} \left(\sum\limits_{e_{ij}\in\E^*} w(e_{ij})\right)$. Typically, the weight of an edge $e_{ij} = \{\x_i,\x_j\}$ is the Euclidean distance between $\x_i$ and $\x_j$ that is $w(e_{ij})=\sqrt{(\x_i - \x_j)\cdot (\x_i - \x_j)'}$. From this fact follows that Euclidean distances are meaningful for the data set, i.e.\,all variables are estimated on the same (or at least a similar) scale. The term ''connected'' refers to the property that there must be a path (i.e.\,a sequence of edges) in $\mathbf{G}$ between any two points of $\X$. It can be proven that the number of edges in $\mathbf{G}$ is always $|\mathbf{E^*}|=n-1$ \citep{Jungnickel2008}. Moreover, it can be shown that the MST is unique if all edges in $\mathbf{E}$ have unique weights. The MST can be efficiently computed even for large data sets. See e.g.\,\cite{Jungnickel2008} for a review of efficient algorithms.

Given the MST, pruning is realized by successively deleting edges in $\mathbf{G}$ according to their length. I.e.\,in the first iteration the longest edge is removed, in the second iteration the second longest edge, and so on. This way $\mathbf{G}$ is split into several subgraphs which are not connected amongst each other. During the pruning process, the subgraphs' cardinality (i.e.\,the corresponding number of observations) declines. The pruning process is stopped if, by deleting the next edge, the cardinality of the largest subgraph would fall below $\lfloor (n+p+1)/2\rfloor$. Stopping at this bound assures that subsequently applied estimators achieve a maximum breakdown point, i.e.\,they are resistant against a maximum level of contamination. The largest subgraph at this point contains more than or exactly $\lfloor (n+p+1)/2\rfloor$ observations and is denoted by $\mathbf{G'} = \left(\X', \E'\right)$ with $\X' \subset \X$ and $\E' \subset \E^*$. This approach was first proposed by \citet{Bennett2001}. A discussion on the robustness properties of the associated estimators can be found in \citet{Kirschstein2013}. Problematically, in most real world data sets much less than $n-\lfloor (n+p+1)/2\rfloor$ outliers occur. To avoid low efficiency of robust estimators, reweighting procedures were proposed to enlarge the uncontaminated subsample.

For MST-based outlier decontamination, reweighting implies that a critical edge length $w_{\alpha}^{crit}$ has to be determined up to which (with a certain probability $\alpha$) the MST consists of uncontaminated observations only. To estimate $w_{\alpha}^{crit}$, a non-parametric approach relying on a finite sample version of Chebychev's inequality is described in \citet{Liebscher2014}. The main idea is to estimate $w_{\alpha}^{crit}$ based on the mean edge length $\mu^w$ and the edge lengths' standard deviation $\sigma^w$. The parameters $\mu^w$ and $\sigma^w$ are estimated based on the edge set $\E'$ of the initial robust subset of $\mathbf{G}'$. For $w_{\alpha}^{crit}$ follows $w_{\alpha}^{crit} = \hat{\mu}^w + \sqrt{\frac{(m^2-1)}{m^2 \cdot (1-\alpha)-m}}\cdot \hat{\sigma}^w$ where $m$ denotes the cardinality of $\E^{'}$. Once $w_{\alpha}^{crit}$ is determined, $\mathbf{G'}$ is ''rebuilt'' by attaching all edges of $\mathbf{G}$ with edge weights smaller or equal to $w_{\alpha}^{crit}$. This way, a still robust but larger subgraph (say $\mathbf{G}''$) is determined whose associated observations are considered as the outlier-free subsample used in further analyses.

\subsection{Production function estimation}

Suppose production can be described by the Cobb-Douglas function:

\begin{equation}
	y_{it}=\alpha^{A}a_{it}+\alpha^{L}l_{it}+\alpha^{K}k_{it}+\alpha^{M}m_{it}+\omega_{it}+\varepsilon_{it},
	\label{eq_1}
\end{equation}

where $y$ is  the natural logarithm of output $Y$, $A$ is land use, $L$  is labour, $K$  fixed capital, $M$  materials (working capital) and $i$ and $t$ are farm and time indices. Lower case letters denote the natural logarithm of the inputs. The $\alpha^{X}$ are parameters to be estimated, and $X \in \left\{A,L,K,M\right\}$ refers to the production factors. $\omega_{it}$ are farm- and time-specific factors known by the farmer but unobserved by the analyst. $\varepsilon_{it}$ are the remaining independent and identically distributed (iid) errors.

First, previous studies on productivity analysis using FADN data focused on this functional form. Flexible functional forms such as the Translog specification are desirable. However, recent studies employing this data set have shown at best mixed results with the Cobb-Douglas specification
\citep{Zhengfei2006,Petrick2013,Latruffe2013}. Furthermore, with increasing degrees of power in polynomials of inputs virtually any data structure can be modelled. This could possibly also mitigate the outlier problematic but it implies adding more and more regressors to the estimating equation. As a result, problems such as multicollinearity are amplified. Therefore, we resort to the specification stated in \eqref{eq_1}.

Next, the $\omega_{it}$ will likely be correlated with the other inputs in \eqref{eq_1} because factor use across farms is usually under control of the farmer. Therefore, the production factors in \eqref{eq_1} are subject to an endogeneity problem. As a result, the OLS estimator which is commonly used as an empirical baseline will produce biased estimates of output elasticities since it neglects the presence of $\omega_{it}$. In empirical practice a typical outcome are upward biased elasticities for variable inputs (e.g. materials). To tackle the endogeneity problem, several strategies have been proposed \citep[cf.][]{Griliches1998}. One route is to assume that $\omega_{it}$ can be clearly separated into individual and time specific fixed effects \citep{Mundlak1961}. However, this approach, so far, has been applied only with mixed success. In order to control for $\omega_{it}$, we apply in this paper a semi-parametric control function approach due to \citet{Wooldridge2009}. This estimator belongs to a class of so-called 'proxy' approaches introduced by \citet{Olley1996} and further developed by \citet{Levinsohn2003}. Control function estimators have already been successfully applied in an agricultural economics context \citep[e.g.][]{Petrick2013, Kloss2014}.These estimators assume that adjustment costs are the main driver of unobserved heterogeneity, $\omega_{it}$. The Wooldridge extension further allows to tackle the collinearity issue \citep{Bond2005}. This problem prevents other estimators (including OLS and traditional fixed effects approaches) from  theoretically identifying the impact of fully variable production factors on productivity.


Assuming the existence of a suitable proxy (e.g. materials) for $\omega_{it}$, we can write:

\begin{equation}
	\omega_{it}=h_t(m_{it}, k_{it}),
	\label{eq_h}
\end{equation}

where $h_t$ is a potentially observable control function and $k_{it}$  is the pre-determined level of capital use at time $t$. It is assumed to evolve according to $k_{it+1}=(1-\delta)k_{it}+inv_{it}$, with $\delta$ as the depreciation rate and $inv$ as investment.

Furthermore, suppose that unobserved productivity follows a first-order Markov process:

\begin{equation}
	\omega_{it}=E[\omega_{it}|\omega_{it-1}]+\xi_{it},
	\label{eq_4}
\end{equation}

where $\xi_{it}$ is an innovation uncorrelated with $k_{it}$, but possibly correlated with the other factors in the production function. Following \citet{Wooldridge2009}, we further assume that $\omega_{it}$ has conditional expectation such that:

\begin{align}
	E[\omega_{it}|k_{it},a_{it-1},l_{it-1},k_{it-1},m_{it-1},\ldots,a_{i1},l_{i1},k_{i1},m_{i1}] \nonumber \\  =E[\omega_{it}|\omega_{it-1}]=g(\omega_{it-1})\equiv g[h(m_{it-1},k_{it-1})],
	\label{eq_7}
	\end{align}
	
where $g$ is an unknown productivity function.

Now, the problem can be formulated by plugging the identity, $\omega_{it}=g[h(m_{it-1},k_{it-1})]$, in \eqref{eq_1} as follows:

\begin{equation}
	y_{it}=\alpha^{A}a_{it}+\alpha^{L}l_{it}+\alpha^{K}k_{it}+\alpha^{M}m_{it}+g[h(m_{it-1},k_{it-1})]+e_{it},
	\label{eq_9}
\end{equation}

where $e_{it}=\xi_{it}+\varepsilon_{it}$. The moment conditions that hold for \eqref{eq_9} are:\footnote{Note, that the orthogonality conditions in \eqref{eq_9} are weakened in a way that only current realisations and one lag of the inputs are assumed to be uncorrelated with the $\varepsilon_{it}$/$e_{it}$.}

\begin{equation}
	E[e_{it}|k_{it},a_{it-1},l_{it-1},k_{it-1},m_{it-1},\ldots,a_{i1},l_{i1},k_{i1},m_{i1}]=0.
	\label{eq_10}
\end{equation}

Hence, in \eqref{eq_9} current and past values of $k$, past values of $a$, $l$ and $m$ and also functions of these may be used as instruments. Given this setup, we can identify the production function parameters by estimating \eqref{eq_9} using instrumental variable estimation with instruments for $a$, $l$ and $m$ \citep[p. 113]{Wooldridge2009}. The function $h$ is approximated by low-order polynomials of first-order lags of $m$ and $k$ while $g$ might be a random walk with drift \citep[p. 114]{Wooldridge2009}.

\subsection{Artificial example}

In order to demonstrate the effects of outliers on non-robust estimations, we discuss a simplified example. Therefore, we simulate an example with 100 farms over 7 periods. The data generating process of these observations is as follows:
\begin{equation}
	y_{it}=0.4\cdot l_{it}+0.6 \cdot k_{it}+\omega_{i}+\varepsilon_{it},
	\label{eq_example1}
\end{equation}
where $y$, $l$ and $k$ are the natural logarithm out output, labour and capital, $\omega_{i}$ represents unobserved heterogeneity with $\omega_{i} \sim N(0,25)$ and $\varepsilon_{it}$ is the remaining disturbance following $N(0,1)$. Labour and capital input are random variables with $N(0, 4)$.

As outliers we generate two data sets with 20 small farms over the same periods with $\omega_{i} \sim N(-5,4)$ and $l_{it},k_{it} \sim N(-5,9)$. This additional data is added to the 'raw' sample. Finally, we arrive at two different outlier contaminated data samples. In the first outlier data set, we set $cor(l_{it}, k_{it}) = 0$ (sample I) and in the second $cor(l_{it} k_{it}) \approx -1$ (sample II), so that labour and capital are almost (perfect) substitutes. This assumption has been chosen for demonstrative purposes. Please note, we further differentiate in this way to illustrate the effects of multicollinearity on outlier infested data and how outlier decontamination is able to mitigate such effects. In both samples the production function for the outlier sets is:
\begin{equation}
	y_{it} = 0.99\cdot l_{it}+0.01 \cdot k_{it} + \omega_{i} + \varepsilon_{it}.
	\label{eq_example2}
\end{equation}
I.e.~in both sets $7 \cdot 20 = 140$ observations are generated by \eqref{eq_example2} which are regarded as outlying from the process assumed in \eqref{eq_example1}.

We present fixed effects regression results to control for unobserved farm specific effects ($\omega_{i}$) for all three data sets applying a) no decontamination, b) univariate outlier decontamination, and c) multivariate decontamination using the pMST method. The estimated production coefficients for labour and capital input as well as final sample size are summarized in \autoref{tab:art_exa_res}.

\begin{table}[h]
\caption{Results of simulated production function example}\label{tab:art_exa_res}
\small

\begin{tabular*}{\textwidth}{L{1.5cm} @{\hskip3ex} ccc @{\hskip8ex} ccc @{\hskip8ex} ccc}
\toprule
decont.  & \multicolumn{3}{c}{raw sample}\hskip8ex  & \multicolumn{3}{c}{sample I}\hskip8ex & \multicolumn{3}{c}{sample II } \\
scheme & est. $l$ & est. $k$ & $N$  & est. $l$ & est. $k$ & $N$  & est. $l$ & est. $k$ & $N$ \\
 \midrule \addlinespace
 with\-out &	0.41$^{***}$ 	& 0.61$^{***}$	&	700 & 0.28$^{***}$	& 0.73$^{***}$	& 840	& 0.22$^{***}$	& 0.80$^{***}$	& 840\\
\addlinespace\hline\addlinespace
uni\-variate & 	0.39"$^{***}$ 	& 0.62$^{***}$	&	526 & 0.24$^{***}$	& 0.74$^{***}$	& 677	& 0.20$^{***}$	& 0.78$^{***}$	& 711\\
\addlinespace\hline\addlinespace
multi\-variate & 	0.40$^{***}$ 	& 0.61$^{***}$	&	498 & 0.42$^{***}$	& 0.62$^{***}$	& 539	& 0.41$^{***}$	& 0.62$^{***}$	& 534\\
\addlinespace\midrule\addlinespace
%
%
\end{tabular*}\\
Notes: *** (**, *) significant at the 1\% (5\%, 10\%) level. The univariate decontamination is based on the average capital productivity per farm. We exclude values outside [Q1-1.5IQR;Q3+1.5IQR]. The multivariate decontamination is based on the pMST procedure.\\
	Source: Authors.
\end{table}

Given an appropriate estimator that controls for unobserved heterogeneity, we are able to recover the 'true' production function parameters, as given in \eqref{eq_example1}, rather precisely for both contaminated samples with the multivariate decontamination method. Univariate decontamination yields no improvement in estimation accuracy.  In fact, results are close to the raw sample. Hence, this simple procedure fails to detect the 'meaningful' outliers in a sense that it cannot detect outliers in dimensions other than the considered one. Therefore, without applying multivariate decontamination, already a relatively small number of outliers (about 16.7 per cent) biases the estimates severely.

Multicollinearity (sample II) increases the problem which results in even more deteriorated estimates. However, multivariate decontamination is able to identify the outliers correctly in this case, too. In particular, this example demonstrates that the presence of multicollinearity does not interfere with the multivariate detection procedure's ability to identify outliers correctly - even in this extreme case of almost perfect correlation between the inputs. Moreover, to put it differently, the pMST procedure might also mitigate multicollinearity. This, as illustrated in this example, works especially well if the outliers are the (major) source of multicollinearity in the data.

In summary, only after controlling for both - unobserved heterogeneity and the effects of outliers - we are capable of obtaining reliable output elasticity estimates. Hence, these two issues have to be treated individually. Given the results of this simple example, an effective outlier decontamination can only be conducted if all model dimensions are considered, i.e. multivariate outlier detection is conducted.

\section{Data}
\label{sec_data}

We use data for field crop farms extracted from the German Farm Accountancy Data Network (FADN) covering the years from 2001 up to 2008. The FADN provides a farm level data set that holds accountancy data for 25 of the 28 EU member states. To represent the heterogeneity of farms and ensure representativeness, a stratified sample is obtained. The stratification criteria are region, economic size and type of farming.  Each year about 80,000 farms are sampled. They represent a population of about 5,000,000 farms in the member states. In each member state a liaison agency is responsible for the data collection and transmission, which consists of about 1,000 variables including structural, economic and financial data.

The farm universe consists of all farms with more than one hectare or those with less than one hectare that provide the market with a specified amount of output. From this universe all non-commercial farms are excluded. To be classified as a commercial farm, a farm must exceed a certain economic size. It is measured in economic size units (ESU). One ESU represents a certain amount of standard gross margin in \euro{} that is periodically adjusted for inflation. In addition, farms are classified by type of farming (TF). To justify the assumption of a homogenous state of technology across farms, we only study field crop farms (TF1) and treat East and West Germany separately in the following analysis because they are structurally distinct. East German agriculture can be characterised by large-scale corporate farms while West Germany is dominated by small- to medium-scale family farms. In addition, after the removal of outliers such a treatment allows carving out similarities and differences more precisely since both German regions are under the same jurisdiction while having historically different forms of agricultural organization. East Germany consists of the five states Mecklenburg-West Pomerania, Brandenburg, Saxony-Anhalt, Thuringia and Saxony. West Germany contains all other states except Berlin and Bremen, which are not represented in the FADN data. From the raw data provided by FADN we constructed panel data sets covering the observed years.

We measure output as the total farm output in \euro{}, labour by the total on-farm hired and family labour working time and land as the utilised agricultural area in ha - including owned and rented  land as well as land in sharecropping.  A persistent issue in estimating production functions has been the specification of the capital variable. Typically, some simple measures of input quantities (such as fertilisers or pesticides) and machinery use (such as fuel expenses or tractor hours) are used in the cross-sectional studies. More sophisticated approaches use inventory methods to estimate real capital service flows by making assumptions about depreciation and capital rental rates \citep{Andersen2011}. In this study, the material or working capital input is proxied by total intermediate consumption in \euro{}. It consists of the total specific costs and overheads arising from production in the accounting year. Among others, it includes cost for fuel, lubricants, water, electricity and seed. We do not include the costs for fertilizer in the materials input as land and fertiliser are highly correlated. This observation implies that these inputs are applied in more or less fixed ratios on the majority of farms. In return, this might induce a multicollinearity problem in the estimations. Nevertheless, due to this correlation, the effect of fertilizer inputs is still captured by the land input even if the former is not included.  Fixed capital inputs are approximated by depreciation of capital assets estimated at replacement value in \euro{}. Such a treatment is consistent with recent literature on production function estimation using firm level data \citep{Olley1996,Blundell2000,Levinsohn2003}. This variable includes depreciation for plantations of permanent crops, buildings and equipment, land improvements, machinery, and forest plantations. Table \ref{tab_vars} summarizes the variable definitions and gives the actual FADN codes.

\begin{table}[!htp]
	\caption{Selection of variables}
	\begin{tabular}{llll}
\cline{1-2}
\textbf{FADN code} & \textbf{Variable description} &  &  \\
\cline{1-2}
\textit{Outputs} &  &  &  \\
SE131 & Total output (\euro{}) &  &  \\
\textit{Inputs} &  &  &  \\
SE011 & Labour input (hours) &  &  \\
SE025 & Total utilised agricultural area (ha) &  &  \\
\pbox{20cm}{F72 + SE300 + \\ SE305 + SE336} &\pbox{20cm}{Costs for seed and seedlings + crop protection + other crop specific \\ costs + overheads (EUR) = materials (\euro{})} &  &  \\
SE360 & Depreciation (\euro{}) = fixed capital &  &  \\
\cline{1-2}
\multicolumn{2}{l}{Source: Authors.}
\end{tabular}
\label{tab_vars}
\end{table}

All monetary values are deflated to real values in 2005 prices using respective price indices. The information was extracted from the Eurostat online database and merged with the country panels. Output was deflated by the agricultural output price index. Fixed capital was deflated by the agricultural input price index for goods and services contributing to agricultural investment, and working capital by the agricultural input price index for goods and services currently consumed in agriculture. \\

\section{Results}
\subsection{Outlier identification}
\label{sec:outid}

\begin{figure}[h!]
\caption{Scatterplot matrix of field crop data with outliers in red}\label{fig_scatter_outl}
  \subfloat[East Germany \label{fig_scatter_dee_outl}]{\includegraphics[width=.48\textwidth]{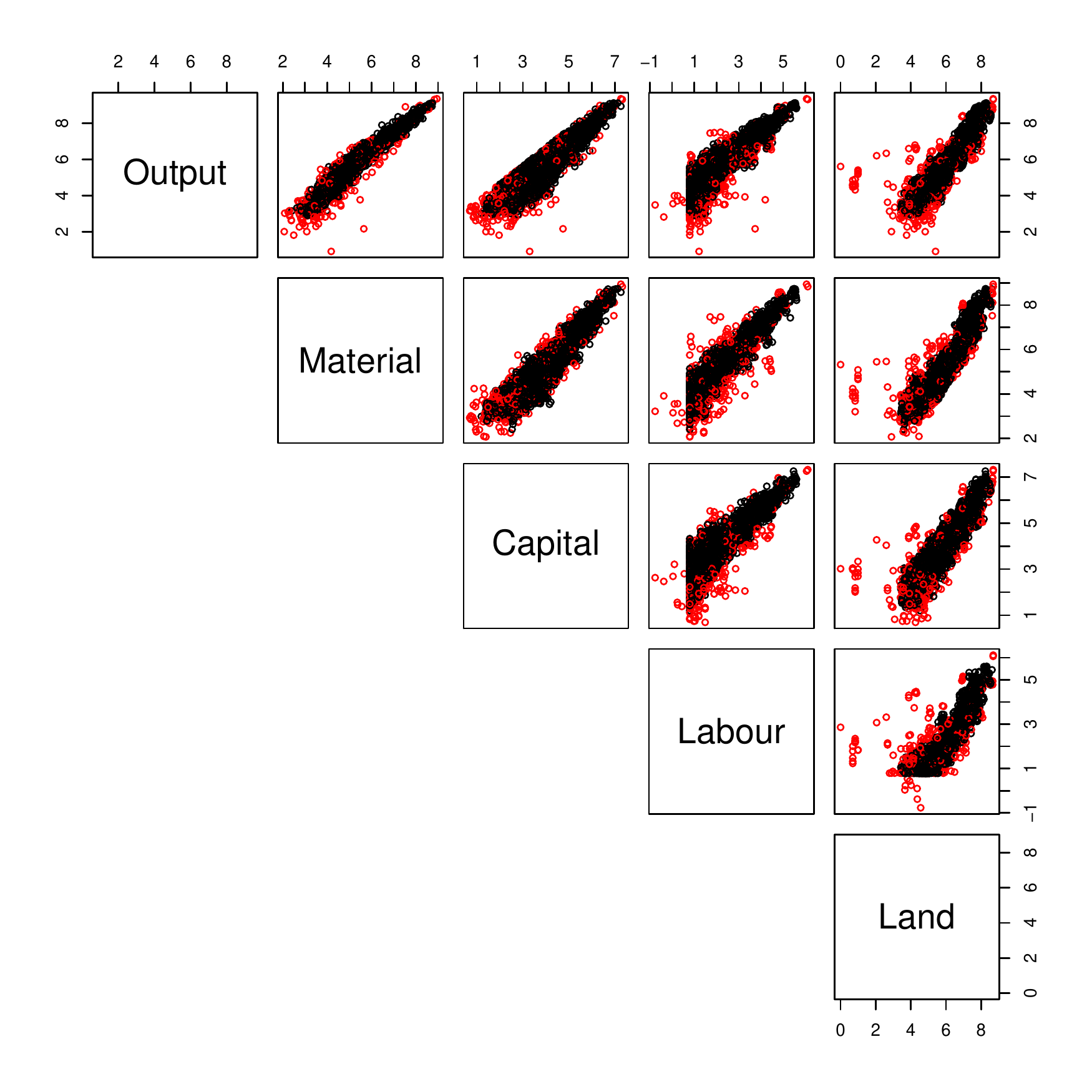}}\hfill
  \subfloat[West Germany \label{fig_scatter_dew_outl}]{\includegraphics[width=.48\textwidth]{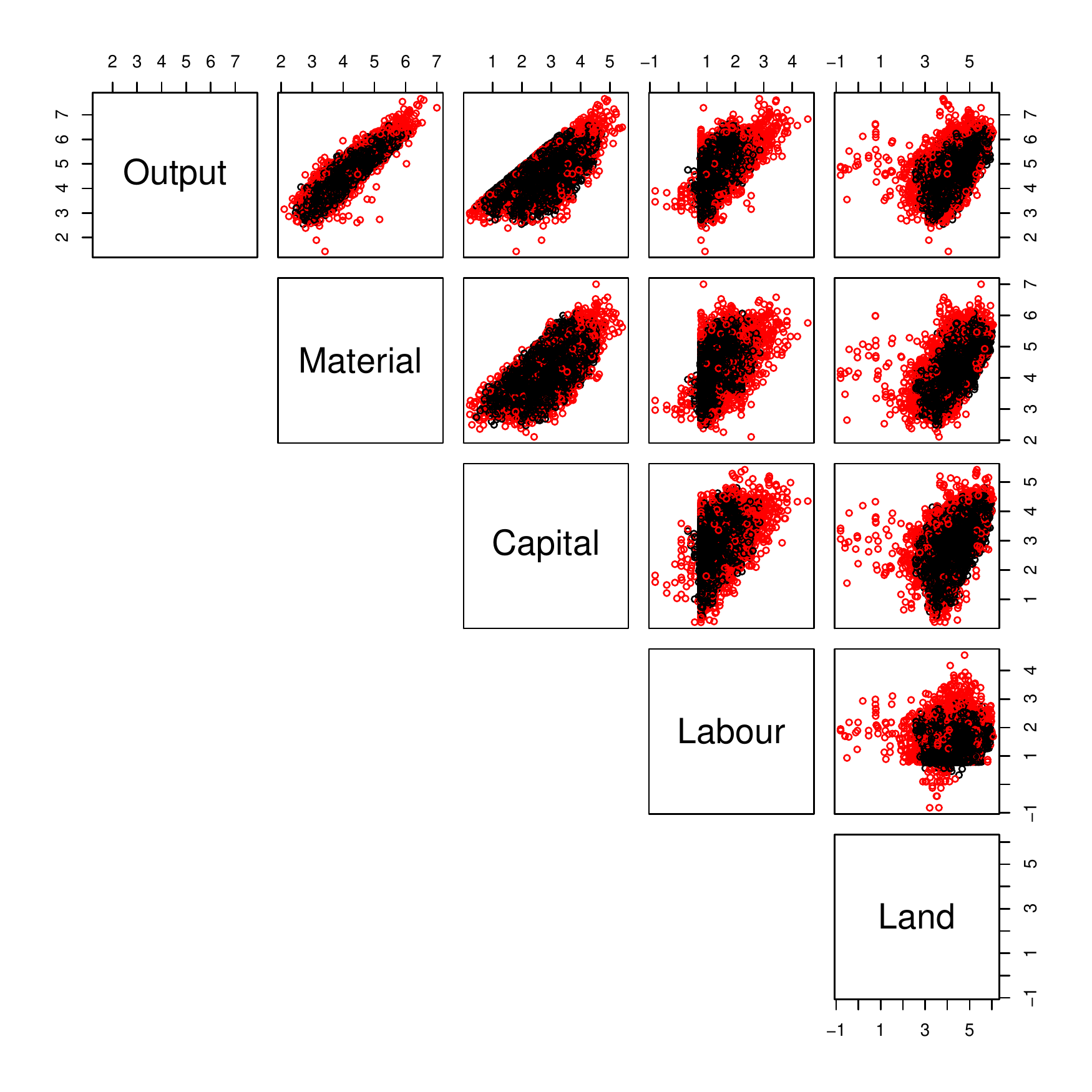}}\\
	Source: Authors.
\end{figure}

The data sets which were finally used for the outlier identification contain a total of 3{,}610 (East Germany) and 8{,}490 observations (West Germany), respectively. Data has been logarithmized to compensate for different scales and heavy tails. Figures \ref{fig_scatter_dee_outl} and \ref{fig_scatter_dew_outl} show the corresponding scatterplot matrices. Neither the plot for East nor for West Germany gives rise to serious concern. In most cases variables show a high pairwise correlation (especially for East Germany). The major part of the data forms a big cluster with some observations scattered around. An interesting effect can be observed for the variable \emph{''Labour''} where the observations seem to be compressed for values somewhere between 0 and 1. This effect results from a substantial amount of ''one-man-companies'' in the data sets which all unanimously stated 2{,}210 hours (i.e., 52 weeks $\cdot$ 5 days $\cdot$ 8.5 hours) as their working input.\\

%

When applying the pMST procedure with $\alpha=0.95$ (as described in Section \ref{sec:out_dec}) on the data, 322 (East) and 1{,}121 (West) potential outliers are identified. Figures \ref{fig_scatter_dee_outl} and \ref{fig_scatter_dew_outl} show the scatterplot matrices with the identified outliers colored in red.\\

%

\noindent The following results are unfolding:
\begin{itemize}
  \item All scattered observations (lying around the main bulk of the data) are identified as outliers (as expected).
  \item It seems that observations lying \emph{within} the main bulk are also identified as outliers, which - at a first glance - might be understood as an undesired result. However, this view leaves the multivariate nature of the data out of consideration, i.e., an observation which is inlying in one scatterplot might be very well an outlier in another bivariate plot or higher dimensional displays.
  \item By tendency, outliers seem to be small farms (with low inputs and low output) as they mainly occur in the lower left corner of the scatterplots (this is visible especially for East Germany in Figure \ref{fig_scatter_dee_outl}).
\end{itemize}

%

\begin{figure}[ht]
	\caption{Characteristics of outliers and non-outliers for field crop data}\label{fig_outliers}
  \subfloat[East Germany\label{fig_outliers_dee}]{\includegraphics[width=.5\textwidth]{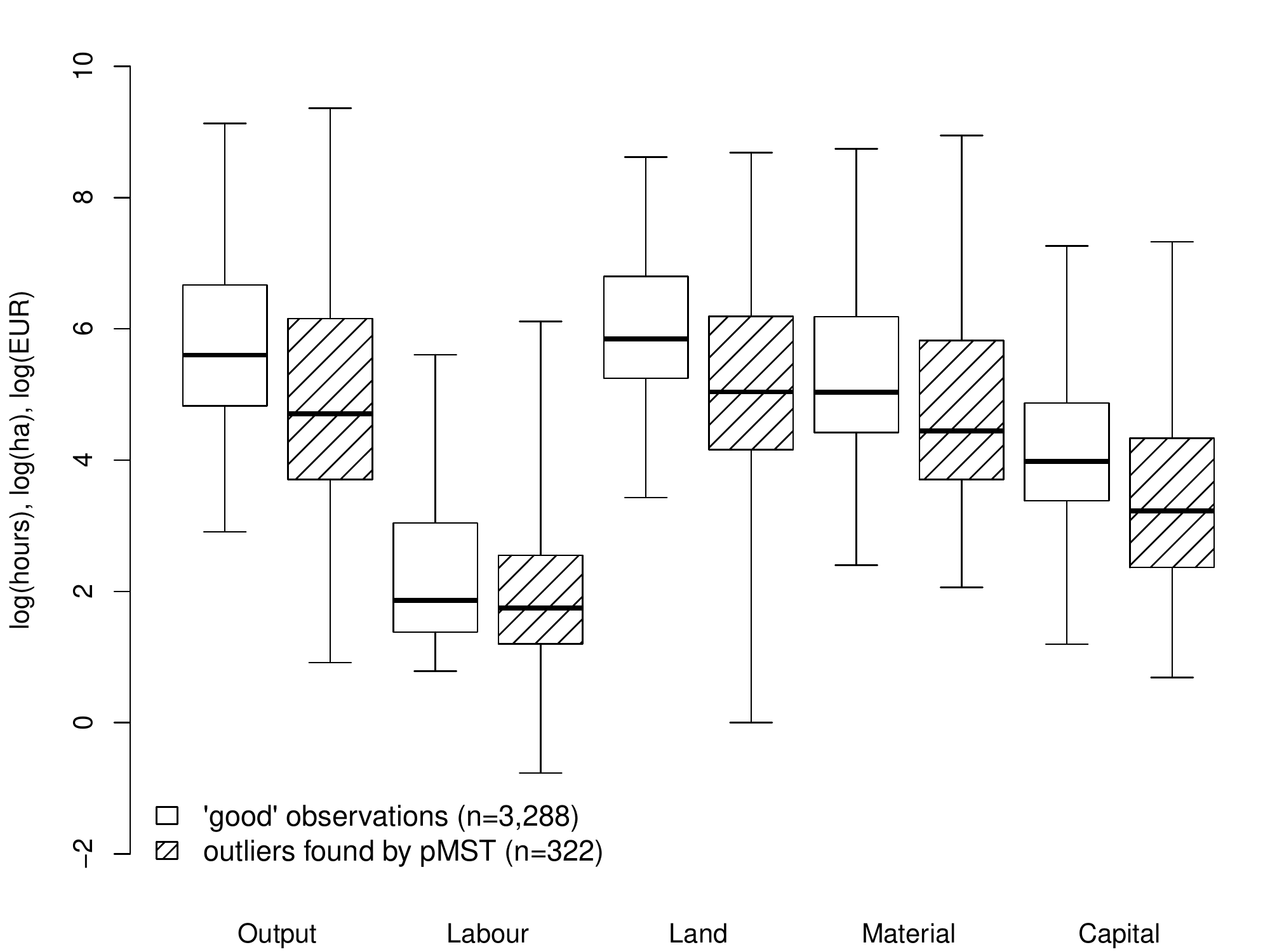}} \hfill
  \subfloat[West Germany\label{fig_outliers_dew}]{\includegraphics[width=.5\textwidth]{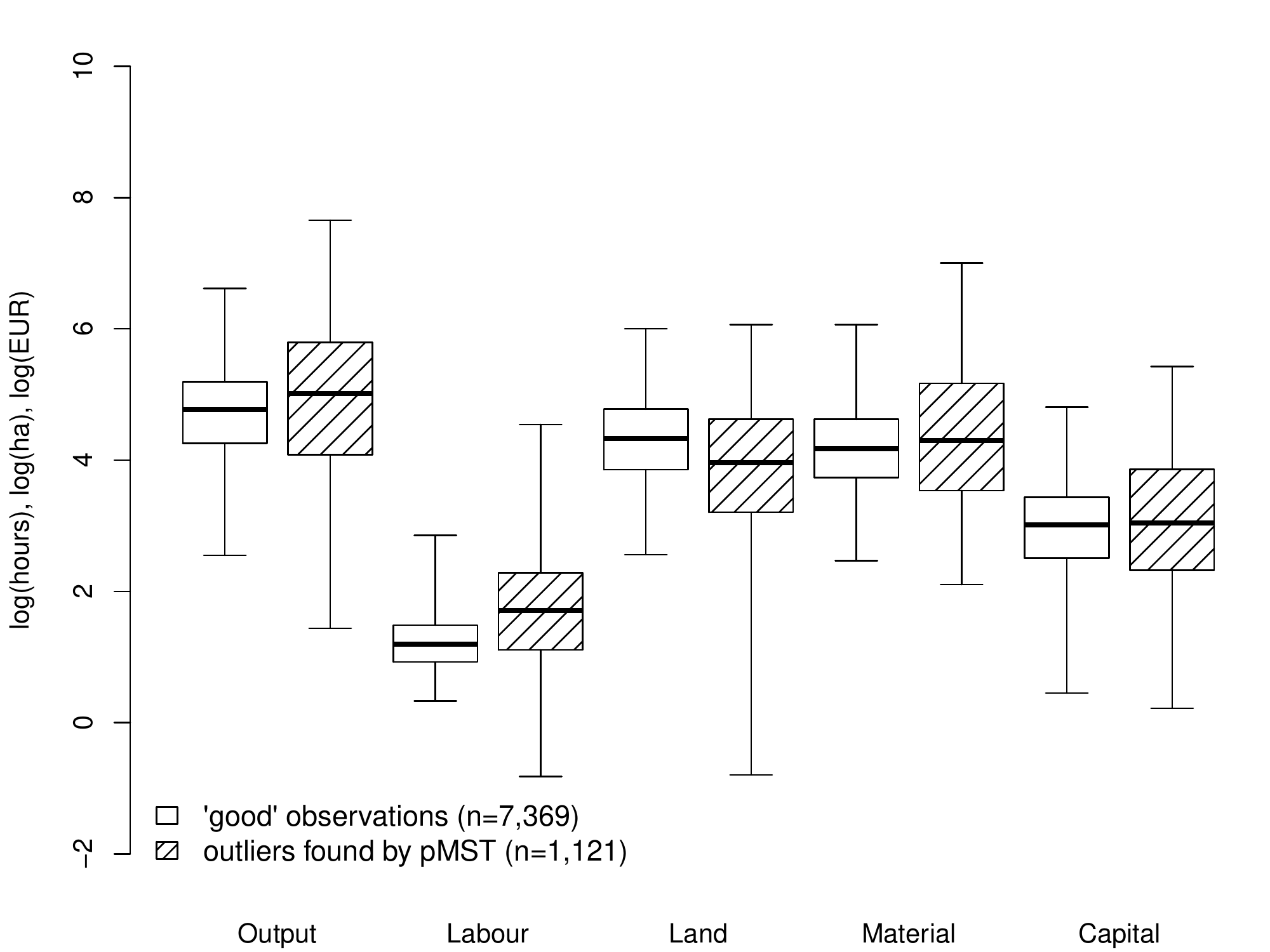}}\\
	Source: Authors.
\end{figure}

The latter point can also be confirmed by having a closer look at the outlier characteristics by means of parallel boxplots; see Figures \ref{fig_outliers_dee} and \ref{fig_outliers_dew}.

In Figure \ref{fig_outliers_dee} it can be seen that the outliers are primarily made up of small companies (presumably of the one-man-companies mentioned earlier) since the boxes and the medians are situated clearly below those of the non-outliers. Additionally, the outliers also cover those observations with very large values in each variable as the (upper) whiskers reach out beyond those of the non-outliers. For the West German data (Figure \ref{fig_outliers_dew}) the situation is not that clear. While the outliers are still made up of the largest and smallest observations in each variable (as expected), the identified outliers seem to be more evenly distributed.

In order to get a better understanding how the outliers divide into ''small'' and ''large'' companies, the approach described in \cite{Liebscher2012a} can be applied: Based on the identified \emph{non}\hyp outliers and their values in the 5 variables a frontier\slash boundary (in concept similar to the Free Disposal Hull, see e.g.\,\citealt{Cooper2007}) can be constructed. Those observations among the \emph{non}\hyp outliers which dominate the remaining non\hyp outliers (in a sense that they possess a higher value in at least one variable while being at least equal in the remaining variables) constitute the upper boundary (see Figure \ref{fig_dom} in the Appendix for an example when considering only the variables Capital and Output of the West German data set). Likewise a lower boundary is constructed (just by switching the sign of the variables). The identified \emph{outliers} can now be assigned to one of the two groups (large\slash small) by examining their position in relation to these boundaries. Outliers lying beyond the upper boundary are considered as ''large'' companies while observations lying below the lower boundary are considered as ''small'' companies. There might also be outliers which lie within the region encapsulated by both boundaries. Table \ref{tab_outl} shows the result of this analysis by giving the number of outliers falling in each region.

\begin{table}[!h]
\caption{Multivariate outliers divided into large and small companies}\label{tab_outl}
\centering
\begin{tabular}{ccccc}
  \hline
   & \parbox[t]{2cm}{\centering small\\ companies} & \parbox[t]{2cm}{\centering large\\ companies} & neither nor & $\sum$ \\\hline
  East & 79 & 6 & 237 & 322 \\
  West & 229 & 236 & 656 & 1121 \\
  \hline
	\multicolumn{2}{l}{Source: Authors.}
\end{tabular}

\end{table}

Note that for these results all five variables have been jointly considered. Therefore the regions are actually hypercuboidal and not planar (as Figure \ref{fig_dom} might suggest). The results generally support the conclusions already drawn from the parallel boxplots. Interestingly, for both East and West Germany, there is a substantial amount of outliers lying within the encapsulated region. These observations, while being identified as outliers, must be located close to the bulk of non\hyp outlying observations. Therefore, if one is interested in a more ''conservative'' approach in outlier identification, one may consider to include these observations in the set of non\hyp outliers again for any follow\hyp on production function analysis.

\subsection{Production function estimation}
\label{sec_pf_results}

We estimate the parameters of the production function by the \cite{Wooldridge2009} (WLP) estimator. It is applied to various samples: a) the full sample without any outlier identification (no-out), b) the "cleaned" subsample resulting after univariate outlier identification (uni-out), c) the "cleaned" subsample resulting after multivariate outlier identification (full-out), and d) the "cleaned" subsample resulting after removing only small and large multivariate outliers (small-large). For case b) observations were dropped if the fixed capital productivity per farm was beyond the upper/lower quartile $\pm$ 1.5 times the interquartile range (IQR). We resort to such a trimming rule because it is prominent in the literature outlined in the introductory section. According to this rule 298 outliers have been detected for East Germany and 1{,}429 for West Germany which differs from the numbers observed in the multivariate case  (see table \ref{tab_outl}). This result is not too surprising as both ways to detect outliers are inherently different. The final estimation samples include farms that have a minimum panel representation of four consecutive years in order to justify the assumption that factor adjustment drives unobserved heterogeneity. This also ensures that the panels do not become "holey" after applying the outlier removal procedure as the detection is done on observations. In addition, it copes well with the assumption that costly factor adjustment drives unobserved heterogeneity \citep{Petrick2013}. In general, our approach ensures that it preserves as much valuable information as possible. For instance, if a firm constantly overreports, it will not be included in the estimation sample.

\subsubsection{Results for East Germany}

Comparing the results in Table \ref{tab_dee_wlp}, it turns out that for the majority of samples the labour coefficient is insignificant and close to zero. Generally, materials is the most important input as it displays output elasticities fluctuating at around $1.0$. Furthermore, the hypothesis of constant returns to scale cannot be rejected for any sample.

\begin{table}[h]
\caption{Results of production function estimation (East Germany, WLP estimator)}\label{tab_dee_wlp}
\small
\begin{tabularx}{1\textwidth}{@{} l @{} X  S[table-format=0.3, table-space-text-post = {***}] S[table-format=0.3] X S[table-format=0.3, table-space-text-post = {***}]S[table-format=0.3] X S[table-format=0.3, table-space-text-post = {***}]S[table-format=0.3] X S[table-format=0.3, table-space-text-post = {***}]S[table-format=0.3] @{}}
	\toprule
	               &  &          \multicolumn{2}{c}{no-out}           &  &          \multicolumn{2}{c}{uni-out}          &  &         \multicolumn{2}{c}{full-out}          &  &        \multicolumn{2}{c}{small-large}        \\
	               &  & {Coeff}       & {SE}                          &  & {Coeff}       & {SE}                          &  & {Coeff}        & {SE}                         &  & {Coeff}       & {SE}                          \\ \midrule
	labour         &  & 0.000         & 0.050                         &  & 0.002         & 0.049                         &  & -0.116$^{***}$ & 0.040                        &  & -0.004        & 0.056                         \\
	land           &  & -0.044        & 0.042                         &  & -0.044        & 0.034                         &  & 0.125$^{***}$  & 0.042                        &  & -0.019        & 0.062                         \\
	materials      &  & 1.015$^{***}$ & 0.136                         &  & 1.034$^{***}$ & 0.158                         &  & 0.884$^{***}$  & 0.147                        &  & 1.001$^{***}$ & 0.145                         \\
	capital        &  & 0.125$^{***}$ & 0.047                         &  & 0.123$^{**}$  & 0.050                         &  & 0.146$^{***}$  & 0.039                        &  & 0.110$^{**}$  & 0.049                         \\ \midrule
	$N$            &  &           \multicolumn{2}{c}{1340}            &  &           \multicolumn{2}{c}{1305}            &  &           \multicolumn{2}{c}{1262}            &  &           \multicolumn{2}{c}{1327}            \\
	Elast of Scale &  & \multicolumn{2}{S[table-format=<0.3]}{1.096}  &  & \multicolumn{2}{S[table-format=<0.3]}{1.115}  &  & \multicolumn{2}{S[table-format=<0.3]}{1.039}  &  & \multicolumn{2}{S[table-format=<0.3]}{1.089}  \\
	p-value CRS    &  & \multicolumn{2}{S[table-format=<0.3]}{0.395}  &  & \multicolumn{2}{S[table-format=<0.3]}{0.381}  &  & \multicolumn{2}{S[table-format=<0.3]}{0.757}  &  & \multicolumn{2}{S[table-format=<0.3]}{0.463}  \\
	p-value Model  &  & \multicolumn{2}{S[table-format=<0.3]}{<0.001} &  & \multicolumn{2}{S[table-format=<0.3]}{<0.001} &  & \multicolumn{2}{S[table-format=<0.3]}{<0.001} &  & \multicolumn{2}{S[table-format=<0.3]}{<0.001} \\
	RSS            &  & \multicolumn{2}{S[table-format=<0.3]}{0.065}  &  & \multicolumn{2}{S[table-format=<0.3]}{0.064}  &  & \multicolumn{2}{S[table-format=<0.3]}{0.051}  &  & \multicolumn{2}{S[table-format=<0.3]}{0.063}  \\ \bottomrule
\end{tabularx}
	Notes: Year dummies included in all models. *** (**, *) significant at the 1\% (5\%, 10\%) level, based on standard errors robust to clustering in groups. CRS: Constant Returns to Scale. RSS (Residual Sum of Squares) is scaled by number of observations and regressors.\\
	Source: Authors.
\end{table}

As expected, results based on the "small-large" subsample and the "uni-out" subsample are quite similar to the "no-out" model. This means estimates and goodness of fit (in terms of residuals' sum of squares, normalized by number of observations and number of parameters to be estimated) are almost equal showing significantly positive effects for materials and capital. Coefficients for labour and land are insignificant and partially negative. A suspicious artefact is an estimated material elasticity greater than 1. Although scale elasticity sums up to values slightly larger than 1, the hypothesis of constant returns to scale cannot be rejected for any model.

Removing the full set of multivariate outliers (full-out subsample) leads to different results. Compared to the other three models, it was possible to identify the land output elasticity and the materials coefficient slightly decreased. Hence, the multivariate outlier decontamination procedure mitigates the multicollinearity problem with regards to these two inputs - a result also observed for West Germany (see below). This observation further implies that a large amount of the multicollinearity due to land and materials is originating from the multivariate outliers.  In addition, farms relying, on average, on a more (fixed) capital intensive production are left in the sample as indicated by the highest capital coefficient among all subsamples. Furthermore, in this particular case results display a higher precision as the standard errors of this parameter is smallest among all subsamples. Likewise, the remaining farms after multivariate decontamination are less working capital intensive.  Moreover, the elasticity of scale is closest to one and the corresponding RSS is the lowest among all models. However, the coefficient of labour is significantly negative for this model which is inconsistent with production theory. Nevertheless, estimation results for the other subsamples, including the more conservatively
multivariately cleaned subsample d), propose a consensus suggesting that labour is an abundant input factor. We observe this issue in the following.

Taking a closer look at the outliers removed additionally compared to the "small-large" subsample reveals that to a large extent the differences between the "small-large" and "full-out" results originate from the observations of a single farm.  A trivariate scatterplot of labour, land, and output shows that the observations of this particular entity are far apart from the remaining data (see the green-coloured points in \autoref{fig_slx} in the Appendix). Removing the  observations of the outlying firm from the "small-large" subsample leads to a significantly negative elasticity estimate for labour (see  \autoref{tab_dee_wlp_slx} in the Appendix). This highlights the influential effect of even a single outlying farm on regression estimates.

In general, removing all (multivariate) outliers leads to a more homogeneous sample. For East Germany this results in an increase in partial correlation between labour and land from 0.81 for the complete data set to 0.90 for "full-out" subsample. The resulting extreme collinearity between both regressors cannot be compensated by the WLP estimator. Therefore, it appears that land and labour may be regarded as a closely intertwined input "package" where one input measures the effect of the other. Omitting either of these two regressors leads to similar estimates as for the fully specified model (compare \autoref{tab_dee_wlp_single} in the Appendix).

To conclude, we find that materials and capital are restrictive inputs for agricultural production in East Germany. The largest influence on production is consistently estimated for the materials input. However, capital shows a significant effect on production in any model. For land and labour the story is more complex due to outliers and a high level of collinearity of both regressors. However, further specifications suggest that labour is an abundant input factor and that the effect of labour and land may be captured by only including the latter.

\subsubsection{Results for West Germany}

The production function estimates for West Germany are summarized in Table \ref{tab_dew_wlp}. In general, the structure of farms in East and West Germany is quite different (as also Figures \ref{fig_outliers_dee} and \ref{fig_outliers_dew} suggest). While in East Germany a lot of very large farms exist (including some farms identified as outlying), in West Germany primarily small and medium scale family farms are prevalent. Hence, one can presume that production technologies in both regions are different to some extent.

\begin{table}[h]
\caption{Results of production function estimation (West Germany, WLP estimator)}\label{tab_dew_wlp}
\small
\begin{tabularx}{1\textwidth}{@{} l @{} X S[table-format=0.3, table-space-text-post = {***}] S[table-format=0.3] X S[table-format=0.3, table-space-text-post = {***}]S[table-format=0.3] X S[table-format=0.3, table-space-text-post = {***}]S[table-format=0.3] X S[table-format=0.3, table-space-text-post = {***}]S[table-format=0.3] @{}}
	\toprule
	               &  &               \multicolumn{2}{c}{no-out}               &  &   \multicolumn{2}{c}{uni-out}   &  &  \multicolumn{2}{c}{full-out}   &  & \multicolumn{2}{c}{small-large} \\
	               &  & {Coeff}                                        & {SE}  &  & {Coeff}                 & {SE}  &  & {Coeff}                 & {SE}  &  & {Coeff}                 & {SE}  \\ \midrule
	labour         &  & 0.186$^{***}$                                  & 0.019 &  & 0.184$^{***}$           & 0.019 &  & 0.167$^{***}$           & 0.023 &  & 0.178$^{***}$           & 0.021 \\
	land           &  & -0.010                                         & 0.016 &  & -0.011                  & 0.016 &  & 0.019                   & 0.020 &  & -0.000                  & 0.018 \\
	materials      &  & 0.802$^{***}$                                  & 0.086 &  & 0.773$^{***}$           & 0.083 &  & 0.803$^{***}$           & 0.091 &  & 0.786$^{***}$           & 0.089 \\
	capital        &  & 0.159$^{***}$                                  & 0.023 &  & 0.169$^{**}$            & 0.023 &  & 0.129$^{***}$           & 0.022 &  & 0.156$^{***}$           & 0.023 \\ \midrule
	$N$            &  &                \multicolumn{2}{c}{3382}                &  &    \multicolumn{2}{c}{3355}     &  &    \multicolumn{2}{c}{3053}     &  &    \multicolumn{2}{c}{3245}     \\
	Elast of Scale &  & \multicolumn{2}{S[table-format=<0.3]}{1.137}  &       & \multicolumn{2}{S[table-format=<0.3]}{1.115}  &       & \multicolumn{2}{S[table-format=<0.3]}{1.118}  &       & \multicolumn{2}{S[table-format=<0.3]}{1.120}    \\
	p-value CRS    &  & \multicolumn{2}{S[table-format=<0.3]}{0.061}  &       & \multicolumn{2}{S[table-format=<0.3]}{0.105}  &       & \multicolumn{2}{S[table-format=<0.3]}{0.135}  &       & \multicolumn{2}{S[table-format=<0.3]}{0.112}    \\
	p-value Model  &  & \multicolumn{2}{S[table-format=<0.3]}{<0.001} &       & \multicolumn{2}{S[table-format=<0.3]}{<0.001} &       & \multicolumn{2}{S[table-format=<0.3]}{<0.001} &       & \multicolumn{2}{S[table-format=<0.3]}{<0.001}   \\
	RSS            &  & \multicolumn{2}{S[table-format=<0.3]}{0.065}  &       & \multicolumn{2}{S[table-format=<0.3]}{0.065}  &       & \multicolumn{2}{S[table-format=<0.3]}{0.060}  &       & \multicolumn{2}{S[table-format=<0.3]}{0.064}    \\ \bottomrule
\end{tabularx}
Notes: Year dummies included in all models. *** (**, *) significant at the 1\% (5\%, 10\%) level, based on standard errors robust to clustering in groups. CRS: Constant Returns to Scale. RSS (Residual Sum of Squares) is scaled by number of observations and regressors. \\
	Source: Authors.
\end{table}

Furthermore, while for all subsamples the assumption of constant returns to scale cannot be rejected, this result is much clearer for estimation samples based on multivariate outlier detection methods. The "no-out" subsample even rejects this assumption at the 10 percent significance level. Orders of magnitude of estimated labour and capital coefficients are smaller in the multivariate compared to the "no-out" and univariate cases - hinting at upward biased coefficients in the latter two. This drives the results more into the direction of returns to scale approaching unity. Additionally, a consistent difference between East and West Germany is that the labour elasticity is significantly positive throughout, indicating that this input is to some extent a scarce factor in West Germany.

Generally, for the case of West Germany, the latest methodological advancements in production function estimation together with multivariate outlier decontamination are able to fully embrace their benefits. This is signalled by - as outlined above - plausible estimates which exhibit positive output elasticities throughout all inputs and, in the case of capital inputs, the highest parameter precision. Finally, RSS is lowest for the full multivariate decontamination scheme.

For West Germany we find similar results for material and capital elasticity as for East Germany. Comparing the "full-out" estimates, coefficients of both variables are quite close for both countries. A similar conclusion can be made for the estimates for the land input. The major difference is the estimate of labour elasticity which is significantly positive in West Germany which may indicate a shortage of labour force compared to East Germany. Additionally, the correlation between labour and land is considerably smaller in West Germany compared to East Germany (about 0.11 to 0.81). This indicates a high level of heterogeneity in the West German data set obviously containing firms with different levels of labour intensity. The opposite is the case for East Germany.

\subsection{Evaluation of Results}

In summary, the two-step approach presented allows for robust and consistent estimation of production functions. By applying a multivariate assessment of outliers, we are able to treat all considered dimensions of agricultural production - resulting in a global assessment of outliers. Perhaps unsurprisingly, the univariate decontamination procedure is not capable in detecting the 'meaningful' outliers. This procedure can only detect conventional outliers beyond some threshold in a single dimension, while the multivariate algorithm at hand may, in addition, also detect such outliers located within the production technology. This is a major advantage compared to the common univariate approach employed. By just dropping observations beyond some threshold the univariate procedure tends to "mis-detect" outliers and potentially drops (keeps) valuable (valueless) information. Results indicate that the pMST procedure indeed detected all scattered observations located around the main bulk of observations as well as the ones within the production technology.

The advantage of the multivariate detection also carries over to the estimation, at least to some extent. It results in improved estimates compared to the results for the other samples, e.g. it allows for a higher precision in the estimation of some parameters. In addition, multivariate detection seems to mitigate the effects of multicollinearity further, however, with better results for West Germany. In both German regions, we observe the lowest RSS for estimation samples based on multivariate decontamination with a clear advantage towards the "full-out" samples. This indicates that the multivariately decontaminated samples also lead to preferable results in terms of goodness of fit. Finally, discarding only extreme multivariate outliers leads to more conservative results closer to the full sample estimates.

\section{Conclusion}

In this paper, we propose a two-step approach for estimating agricultural production functions robustly. In the first step, a non-parametric multivariate decontamination procedure based on pruning the minimum spanning tree of a data set is used to determine an outlier-free subsample. In contrast to univariate extreme value detection which is the prevalent approach in empirical economics practice, this method assesses outliers considering all observed dimensions of agricultural production. In the second step, production functions are estimated using the recently proposed panel data estimator by \cite{Wooldridge2009} to mitigate endogeneity as well as collinearity issues. The WLP estimator promises consistent estimates by addressing endogeneity and collinearity problems. In our simulated example, we show that given an appropriate estimator only multivariate decontamination leads to estimates close to the true parameter values.

In the empirical section, we apply the two-step procedure to East and West German agricultural field crop data from the FADN data base. The analysis reveals that many outlying observations are made up of relatively small farms. In addition, outliers were also detected within the main bulk of observations, which cannot be detected with a univariate approach. Hence, future empirical analyses utilising FADN data should apply a multivariate outlier decontamination procedure. The estimated production functions show that the WLP estimator delivers convincing results with respect to input elasticities and returns to scale when it is applied to outlier-free subsamples derived by the multivariate pMST approach. The estimates for East and West Germany show similarly positive elasticities for capital (working and fixed) as well as scale elasticities close to one. By moving from the univariate to the multivariate decontaminated sample a picture unfolds in which, on average, more fixed capital and less working capital farms remain in the East German sample; for the West German one it is the other way around. A key advantage of the multivariate decontamination is that it potentially helps to further mitigate multicollinearity problems (see simulated example). Generally, estimations based on this sample provide parameter estimates of previously unidentified coefficients (land in East Germany) or a higher precision in estimation. In case of East Germany, this does come at some cost as we observe a negative and significant labour coefficient.  Estimates of labour and land elasticities have to be interpreted carefully due to a  high correlation of land and labour.  However, results for the other specifications and subsamples, including the more conservative multivariate solution, suggest that labour is no scarce factor in East Germany while land to some extent is. 

In essence, the proposed two-step approach reveals that production technology in East and West Germany is not that different as it seems at the first glance. This is even more important as West Germany shows a more heterogeneous set of firms compared to East Germany. Nonetheless, in both regions farms rely on material and capital inputs in a very similar way.
	
The most remarkable remaining difference between East and West Germany is the labour elasticity, which is significantly positive for West Germany. This can be interpreted as an indication of a shortage of labour force in West Germany whereas East Germany does not suffer from such a restriction. Similarly, for both regions fixed capital constitutes an equally restrictive input factor whereby the materials input is the far most important input.  Therefore, policy reforms should aim to ease the access of agricultural companies to capital and labour force, particularly in West Germany.

Statistical outlier analysis in general has also its limitations. As it is only a statistical tool it will not absolve the researcher from thinking about the scope of research. With regard to production function estimation, statistical outlier analysis is a mean to homogenize production technologies. Therefore, the statements made in this paper are with respect to the prevalent production technology in both German regions. However, many of the outliers detected are small farms. If such farms are of interest to the research question at hand, the outlier analysis, uni- or multivariate, needs to be reconsidered.


\pagebreak
\section*{Appendix}

\begin{figure}[h!]
\caption{West German field crop data}\label{fig_dom}
  \includegraphics[width=.6\textwidth]{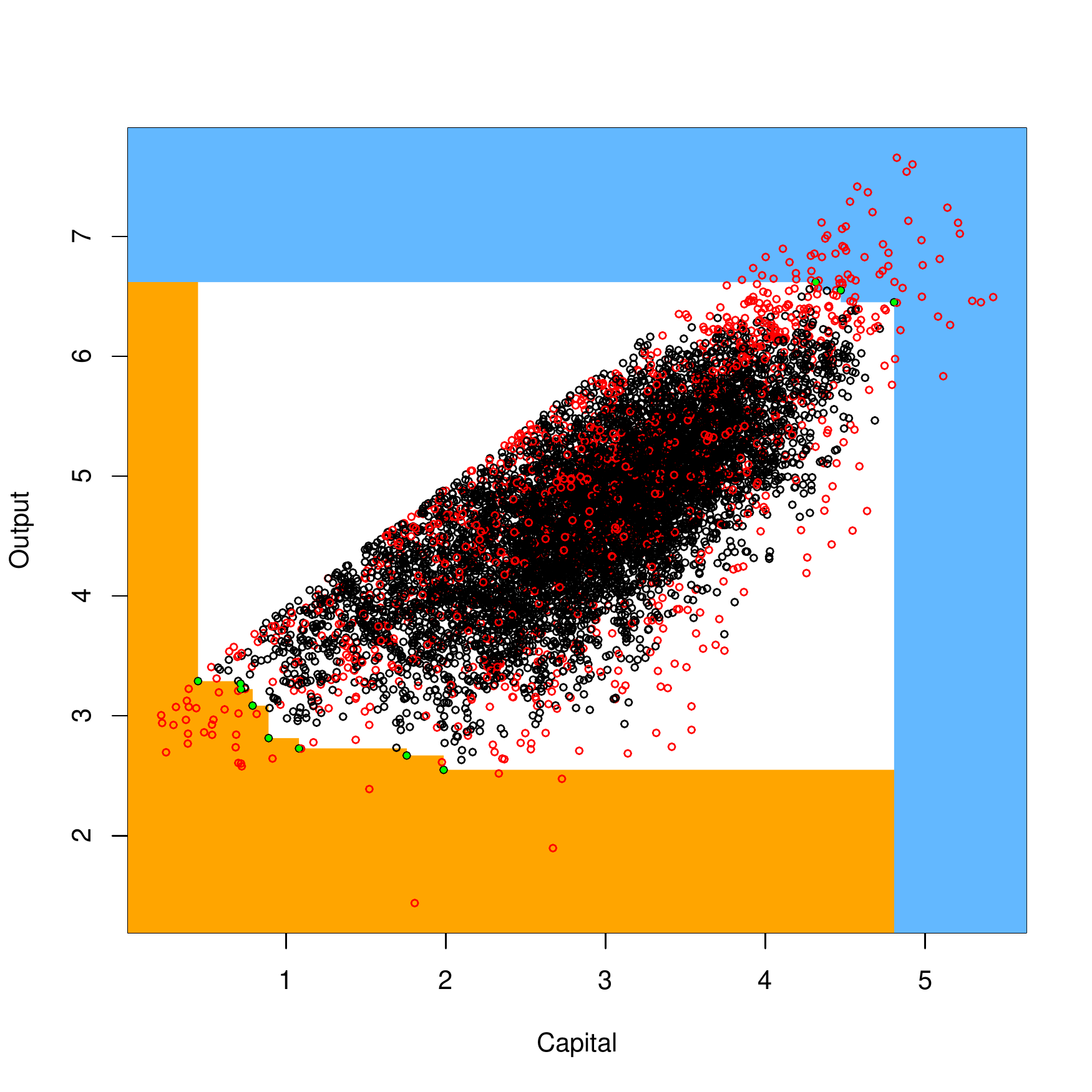}\\
  Notes: Boundaries (green dots) given by the identified non\hyp outliers (black dots). Outliers (red dots) lying in the blue region are considered as large companies and those lying in the orange region as small companies, respectively.\\
	Source: Authors.
\end{figure}

\begin{figure}[h!]
\caption{3D scatterplot of East German data ("small-large" subsample) }\label{fig_slx}
 \includegraphics[width=.8\textwidth]{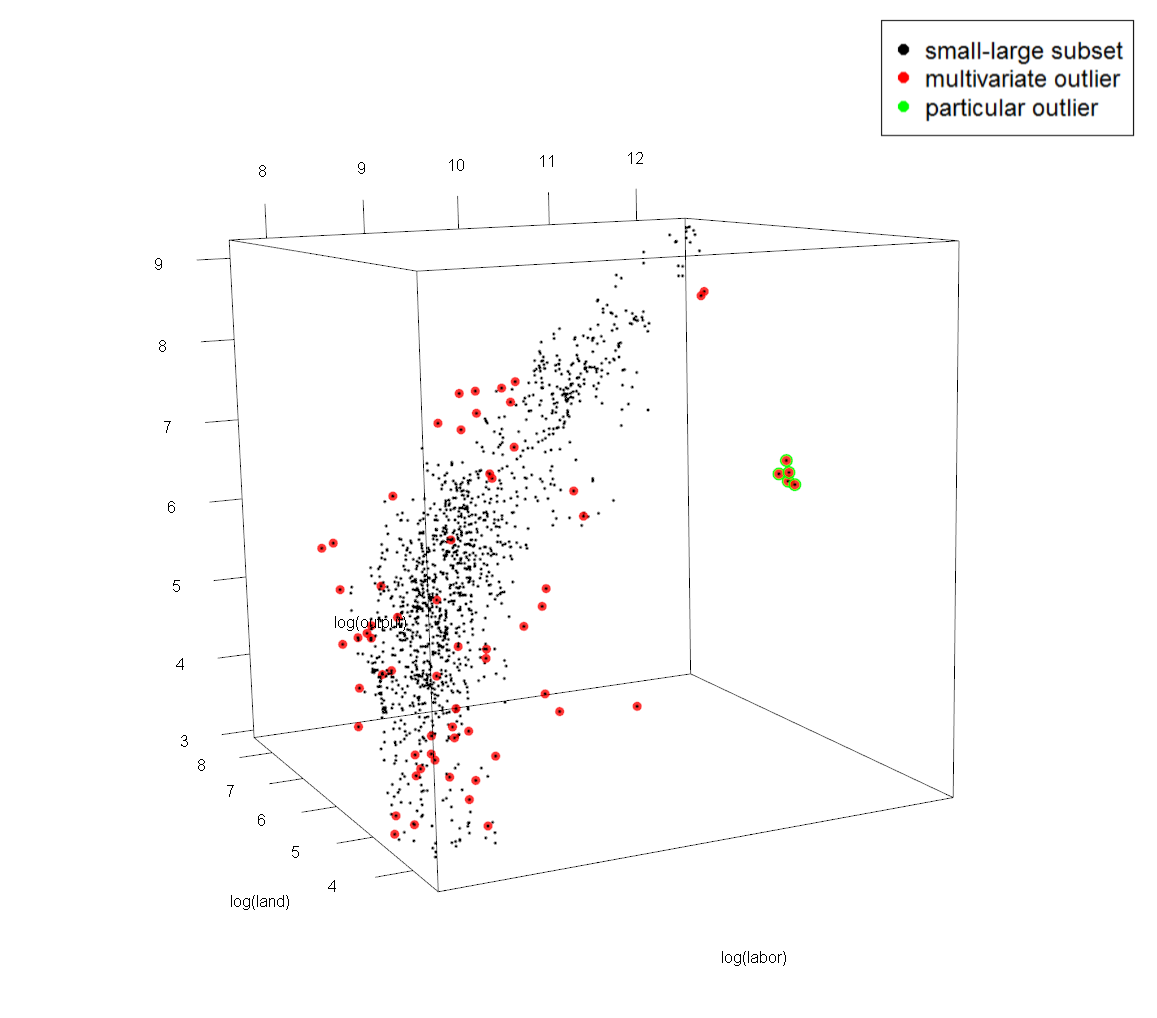}\\
Notes: Dimensions: output, land and labour. Black dots denote the "small-large" subsample (all points). Points encircled with red constitute multivariate outliers. Red dots with green edges denote the outlying observations of interest.\\
	Source: Authors.
\end{figure}

\begin{table}[h]
\caption{Results of production function estimation  ("adjusted" small-large sample, East Germany)}\label{tab_dee_wlp_slx}
\small
\centering
\begin{tabularx}{0.42\textwidth}{l X S[table-format=0.3, table-space-text-post = {***}] S[table-format=0.3]}
	\toprule
	               &  &   \multicolumn{2}{c}{adjusted small-large}    \\
	               &  & {Coeff}       & {SE}                          \\ \midrule
	labour         &  & -0.074$^{*}$  & 0.039                         \\
	land           &  & 0.059         & 0.041                         \\
	materials      &  & 0.969$^{***}$ & 0.143                         \\
	capital        &  & 0.105$^{**}$  & 0.049                         \\ \midrule
	$N$            &  &           \multicolumn{2}{c}{1322}            \\
	Elast of Scale &  & \multicolumn{2}{S[table-format=<0.3]}{1.059}  \\
	p-value CRS    &  & \multicolumn{2}{S[table-format=<0.3]}{0.619}  \\
	p-value Model  &  & \multicolumn{2}{S[table-format=<0.3]}{<0.001} \\
	RSS            &  & \multicolumn{2}{S[table-format=<0.3]}{0.060}  \\ \bottomrule
\end{tabularx} \\
\raggedright	Notes: Year dummies included in all models. *** (**, *) significant at  the 1\% (5\%, 10\%) level, based on standard errors robust to clustering in groups. RSS (Residual Sum of Squares) is scaled by number of observations and regressors. \\
	Source: Authors.
\end{table}

\begin{table}[h]
\caption{Results of regressions leaving labour and land out (full-out sample, East Germany)}\label{tab_dee_wlp_single}
\small
\centering
\begin{tabularx}{0.6\textwidth}{l X S[table-format=0.3, table-space-text-post = {***}] S[table-format=0.3] X S[table-format=0.3, table-space-text-post = {***}] S[table-format=0.3] }
	\toprule
	              &  & \multicolumn{2}{c}{full-out} &  & \multicolumn{2}{c}{full-out}   \\
	              &  & {Coeff}         & {SE}         &  & {Coeff}         & {SE}           \\ \midrule
	labour        &  & {--}            & {--}         &  & -0.086$^{**}$ & 0.037        \\
	land          &  & 0.097$^{**}$  & 0.041      &  & {--}            & {--}           \\
	materials     &  & 0.813$^{***}$ & 0.146      &  & 0.969$^{***}$ & 0.143        \\
	capital       &  & 0.140$^{***}$ & 0.038      &  & 0.150$^{***}$ & 0.040        \\ \midrule
	$N$           &  &  \multicolumn{2}{c}{1262}  &  &  \multicolumn{2}{c}{1262}    \\
	p-value Model &  & \multicolumn{2}{S[table-format=<0.3]}{<0.001} &  & \multicolumn{2}{S[table-format=<0.3]}{<0.001}   \\
	RSS           &  & \multicolumn{2}{S[table-format=<0.3]}{0.052}  &  & \multicolumn{2}{S[table-format=<0.3]}{0.054}    \\ \bottomrule
\end{tabularx} \\
\raggedright	Notes: Year dummies included. *** (**, *) significant at the 1\% (5\%, 10\%) level, based on standard errors robust to clustering in groups. RSS (Residual Sum of Squares) is scaled by number of observations and regressors. \\
	Source: Authors.
\end{table}



\end{document}